\documentclass[runningheads]{llncs}
\usepackage[T1]{fontenc}
\usepackage{graphicx}
\usepackage{url}
\usepackage{color}

\usepackage{hyperref}
\hypersetup{
    colorlinks=true,
    linkcolor=blue,
    filecolor=blue,      
    urlcolor=blue,
    citecolor=blue,
}

\usepackage{caption} 
\usepackage{listings}
\usepackage{float}
\usepackage{appendix}

\usepackage{tabularx}
\usepackage{wrapfig}
\usepackage[ruled,vlined]{algorithm2e}
\usepackage{caption}
\usepackage{marvosym}

\usepackage{bbding}
\begin{document}

\title{An Agile Formal Specification Language Design Based on K Framework}
%
%

\author{Jianyu Zhang\inst{1} \and
Long Zhang\inst{2(}\textsuperscript{\normalsize{\Letter}}\inst{)} \and
Yixuan Wu\inst{3} \and
Feng Yang\inst{2} 
}
\authorrunning{F. Author et al.}
%
\institute{School of Automation Engineering, University of Electronic Science and Technology of China, Chengdu, China \and
National Key Laboratory of Science and Technology on Information System Security, AMS, Beijing, China\\
\email{zhanglong10@nudt.edu.cn}
\and
School of Electronics and Information, Northwestern Polytechnical University, Xi’an, China \\}
\maketitle             

\begin{abstract}

Formal Methods (FMs) are currently essential for verifying the safety and reliability of software systems. However, the specification writing in formal methods tends to be complex and challenging to learn, requiring familiarity with various intricate formal specification languages and verification technologies. In response to the increasing complexity of software frameworks, existing specification writing methods fall short in meeting agility requirements. To address this, this paper introduces an Agile Formal Specification Language (ASL). The ASL is defined based on the K Framework and YAML Ain't Markup Language (YAML). The design of ASL incorporates agile design principles, making the writing of formal specifications simpler, more efficient, and scalable. Additionally, a specification translation algorithm is developed, capable of converting ASL into K formal specification language that can be executed for verification. Experimental evaluations demonstrate that the proposed method significantly reduces the code size needed for specification writing, enhancing agility in formal specification writing. 

\end{abstract}
\section{Introduction}\label{sec1}
As the code framework of modern software becomes more complex, its security and reliability are becoming a research concern. Existing methods such as strict software development processes and various testing methods have been adopted to ensure high reliability of software \cite{mayeda2021evaluating,tauqeer2021analysis}. However, these traditional security assurance methods cannot meet the demand for simultaneous improvement of reliability and agility \cite{kuhrmann2021makes}. These methods are also difficult to completely prove the correctness of the system in theory \cite{mukherjee2020defining}.

Formal methods diverge from traditional security assurance approaches. They employ rigorous mathematical techniques, including formal logic and mathematical reasoning, to verify the correctness of software systems against specified properties and specifications. Formal methods are becoming an important means of system security assurance and are widely used \cite{kulik2022survey}.

Although formal methods can effectively prove the security of a target system, writing formal specifications is intricate. This process demands developers to possess not only system development expertise but also proficiency in mathematical logic. Furthermore, a thorough understanding of diverse formal specification languages and verification technologies is essential \cite{gleirscher2023manifesto,garavel20202020,gleirscher2020formal,krichen2023improving}. These requirements, coupled with the frequent lack of integration with standard integrated development environments \cite{hahnle2019deductive}, challenge the alignment of formal methods with the agility that is increasingly crucial in software development \cite{tsilionis2024scaling}.

Therefore, it is imperative to explore more agile solutions under the premise of high reliability. This paper aims to develop a paradigm of formal specification language that incorporates agile design principles \cite{fowler2001agile}, with the goal of reducing the complexity of specification writing in formal verification. Agile design concepts and techniques within the field of software engineering emphasize simplicity, efficiency, flexibility, and continuous feedback to enhance the quality and adaptability of software development \cite{spagnoletti2021agile}.

\noindent\textbf{Challenges. }From existing work, the conflict between formal methods and the requirements of agility presents a core challenge for this paper, as illustrated by the following points:

\begin{itemize}
    \item To support the description of complex properties and systems, formal methods require a rigorous logical foundation and formal language \cite{gleirscher2023manifesto}. This requirement can make their meanings and formal expressions obscure. As a result, it may lead to difficulties in writing and understanding formal specifications. In contrast, agility demands concise language and intuitive expression \cite{al2020agile}.
    \item To facilitate formal verification, formal methods often necessitate working within a unified logical system. Each formal method typically has its unique specification language syntax, which is often incompatible with specifications from other methods. Agility, however, requires the use of familiar, efficient, and convenient methods and tools \cite{tsilionis2024scaling}.
\end{itemize}

\begin{figure}
    \centering
    \includegraphics[width=\textwidth]{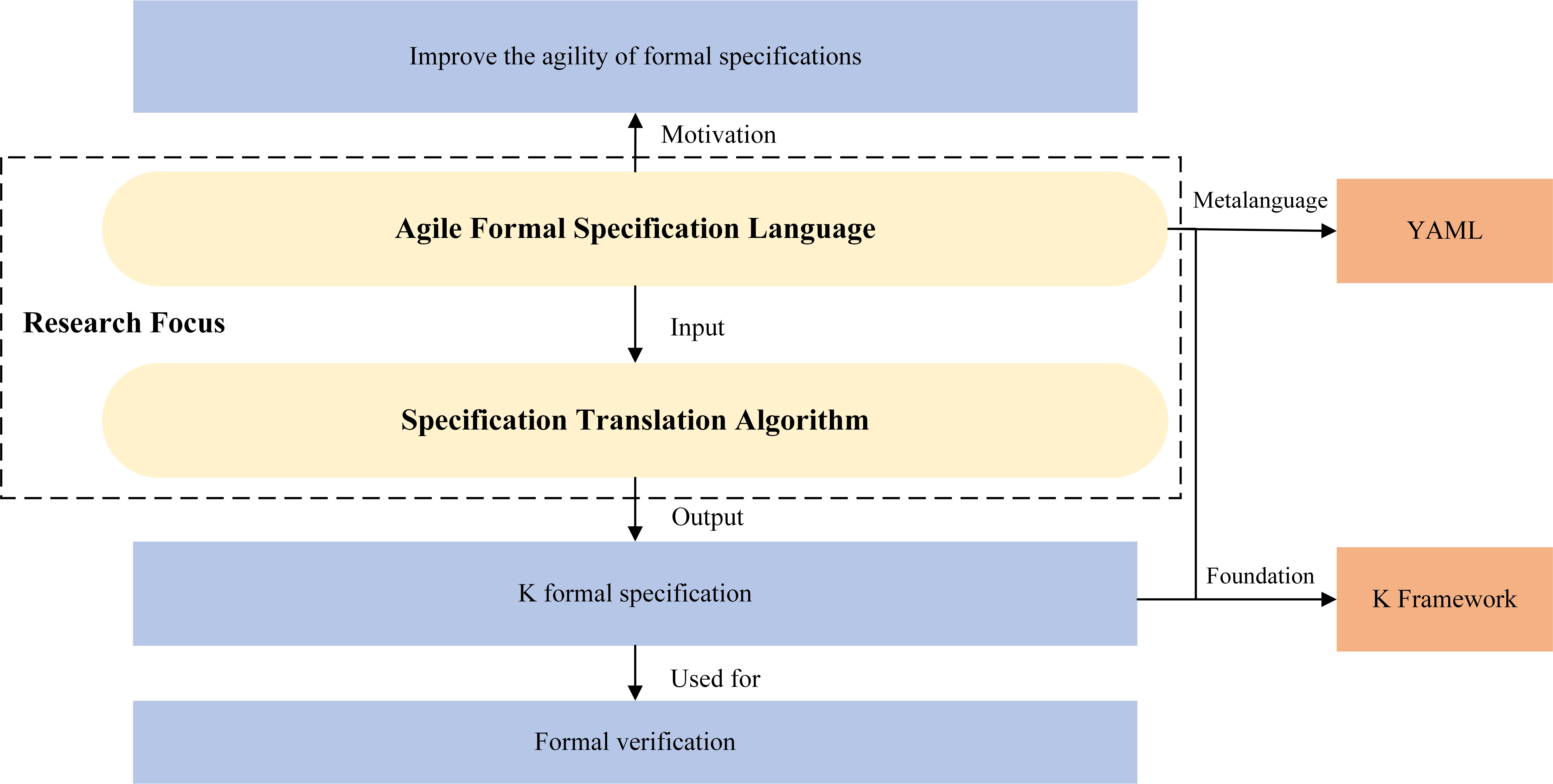}
    \caption{Overall technical framework for this work.}
    \label{fig1}
\end{figure}

\noindent\textbf{Overviews. }To address the aforementioned challenges, the core idea of this research is the proposal of a formal specification language. This language is founded on the K Framework \cite{roșu2010overview,rosu2017k,chen2020technical}, utilizing a structured and parse-friendly YAML \cite{ben2009yaml} as a metalanguage. It is designed to integrate fine-grained specification agility needs, aiming to be simple, efficient, and scalable\cite{fowler2001agile}. Additionally, this research develops a translation algorithm. This algorithm converts agile formal specifications into K formal specifications. The overall technical framework is illustrated in Figure. \ref{fig1}.

K Framework is an executable semantics framework with language independence, capable of automatically generating verification tools based on formal language specifications. The K Framework is based on Matching Logic \cite{rocsu2020matching}, a variant of first-order logic. Matching Logic is used for describing and reasoning about structures. It is known for its concise and intuitive expression capabilities. These attributes are crucial for the K Framework’s ability to describe numerous complex programming languages \cite{chen2021matching}. The powerful logical expression ability and scalability of the K Framework make it chosen as the basis of ASL in this paper.

YAML is a lightweight, easy-to-read, and write data serialization language, widely used in configuration files and data interchange formats \cite{ben2009yaml}. Despite its less efficient performance in certain aspects compared to JavaScript Object Notation(JSON) \cite{bray2014javascript}, YAML has readability, ease of use, and support for multiple tools. These features align well with the agile philosophy. Therefore, YAML was selected as the metalanguage for ASL.

\noindent\textbf{Contributions. }The core contributions of this paper are as follows: 

\vspace{-0.5cm}

\begin{itemize}
    \item In response to the complexity and lack of agility in writing formal specification languages, an Agile Formal Specification Language is proposed. This language enables the completion of formal specification writing through concise and readable syntax.
    \item To facilitate verification of agile formal specifications, a specification translation algorithm has been developed. This algorithm transforms agile formal specifications into existing verifiable formal specifications.
\end{itemize}

\vspace{-0.3cm}

The structure of the remainder of this paper is as follows: Section \ref{sec2} introduces the technical background and motivation of this paper. Section \ref{sec3} describes the design of the proposed Agile Formal Specification Language, elucidating its realization of agility principles. Section \ref{sec4} delineates the specification translation algorithm. Section \ref{sec5} employs experimental validation to assess the method’s efficacy in enhancing specification agility. Concluding remarks and directions for future research are provided in Section \ref{sec6}.

\section{Background and Motivation}\label{sec2}
This paper primarily focuses on the intersection of two research domains: formal verification methods and tools and agile software development approaches. Consequently, this section reviews work within these realms to establish a comprehensive foundation for the study and introduces the motivation behind this paper.

\noindent\textbf{Formal Verification Methods and Tools}. To ensure the security and reliability of systems, academia and industry continuously explore more efficient and convenient formal methods and tools. Rosu and Chen et al. \cite{rosu2017k,chen2020technical} introduced the K Framework based on Matching Logic \cite{chen2021matching,rocsu2020matching}, capable of generating verifier tools from formal language definitions. It has been utilized to define complete executable formal semantics for numerous real-world languages. Stefuanescu et al. \cite{stefuanescu2016semantics} instantiated the K Framework with the semantics of C, Java, and JavaScript. The generated verifiers successfully checked the functional correctness of programs in these languages. Kroening et al. \cite{kroening2014cbmc} proposed CBMC, a bounded model checker for C programs. CBMC translates input C programs into a formula and employs a Boolean Satisfiability Solver (SAT) \cite{fichte2023silent} to determine the presence of executions that violate specifications. It has seen widespread use in C program verification. Heizmann and Dietsch et al. \cite{dietsch2023ultimate} developed Ultimate Automizer, an automated program verification tool that employs trace abstraction algorithms and nested interpolators for computation in program analysis. This tool has achieved notable success in the Software Verification Competition (SV-COMP) \cite{beyer2023competition}. Additionally, there are widely-used formal verification methods and tools such as Z3 Theorem Prover \cite{de2008z3}, Isabelle Theorem Prover \cite{paulson1994isabelle}, Coq Proof Assistant \cite{bertot2013interactive}, and Simple Promela Interpreter (SPIN) \cite{holzmann2004spin}.

The aforementioned research has enhanced the efficiency of formal verification and promoted its application. However, the issue of complex specification writing remains. As modern software architectures become increasingly intricate, they fail to meet the growing demand for agility. There is a need to develop a formal specification language that possesses agile attributes.

\noindent\textbf{Agile Software Development Approaches}. Agile, as a software development methodology, emphasizes adaptability to change and the continuous delivery of high-value software \cite{spagnoletti2021agile}. Fowler et al. \cite{fowler2001agile} introduced the Agile Manifesto. The Agile Manifesto underscores the principle of simplicity by aiming to maximize the reduction of unnecessary work as a core tenet of agile methods \cite{daraojimba2024comprehensive}. Subsequent research on agile development methodologies has largely been guided by the principles outlined in the Agile Manifesto \cite{kaushik2016dsdm}. Schwaber et al. \cite{schwaber1997scrum} proposed the Scrum Development Process (SCRUM). They treated the system development process as a controlled black box, emphasizing flexibility and responsiveness to changing requirements. Beck et al. \cite{beck1999embracing} advocated for Extreme Programming (XP). XP achieves continuous transformation in software development through iterative cycles that include planning, analysis, and design activities.

The research on agile development principles and methods has significantly enhanced the efficiency and quality of software development, making it more streamlined and convenient. However, in the realm of formal verification, there is a lack of corresponding agile methodologies or tools. Developers are still required to manually write extensive and complex formal specifications. 

\noindent\textbf{Motivation}. To address the aforementioned issue, the goal of this paper is to integrate agile development principles into the design of formal specification languages, thereby addressing the complexity involved in writing formal specifications. Consequently, we have established the following design objectives for the specification language we propose to develop:

\begin{itemize}
    \item \textbf{Specification understanding agility}: The specification language should be simple and easy to read, possessing structured features, and adhere to the principle of Occam's Razor \cite{mcfadden2023razor} by retaining only the necessary core concepts required for formal specifications.
    \item \textbf{Specification writing agility}: The specification language should support the reuse or extension of various definitions and specifications for functions and types, thereby avoiding redundant writing.
    \item \textbf{Specification extension agility}: The specification language should be easily scalable and modifiable in terms of syntax, allowing formal specifications to support the verification of different target languages.
\end{itemize}

\section{Agile Formal Specification Language Design}\label{sec3}
This section details the syntax and language definition of the proposed Agile Formal Specification Language, and explains its agility.
\subsection{Specification Language Syntax Composition}
The syntax of agile formal specifications is composed of three main parts: (1) the YAML metalanguage, (2) the ASL keywords, and (3) the expression syntax.

Firstly, YAML is used as the metalanguage. This means that the syntax of the original YAML is adopted. Consequently, all valid agile formal specifications are valid YAML files. This allows ASL to leverage the language features and libraries of YAML.

Secondly, the core concepts of the Agile Formal Specification Language are types, functions, and specifications. Types are employed to define data types. Functions are used to describe system functionalities. Specifications are assertions about system behaviors. They restrict various system behaviors in a concise and intuitive manner.

Lastly, in the expression syntax, the expressions of agile formal specifications follow the syntax of the K Framework \cite{rosu2017k}. For more details, please refer to the official documentation of the K Framework (\url{https://kframework.org/}).

\subsection{Specification Language Definition}
This section delineates the definition of ASL. To visually depict the method of writing specifications, this paper employs a combination of figures and text for explanation.

\begin{figure}
    \centering
    \includegraphics[width=\textwidth]{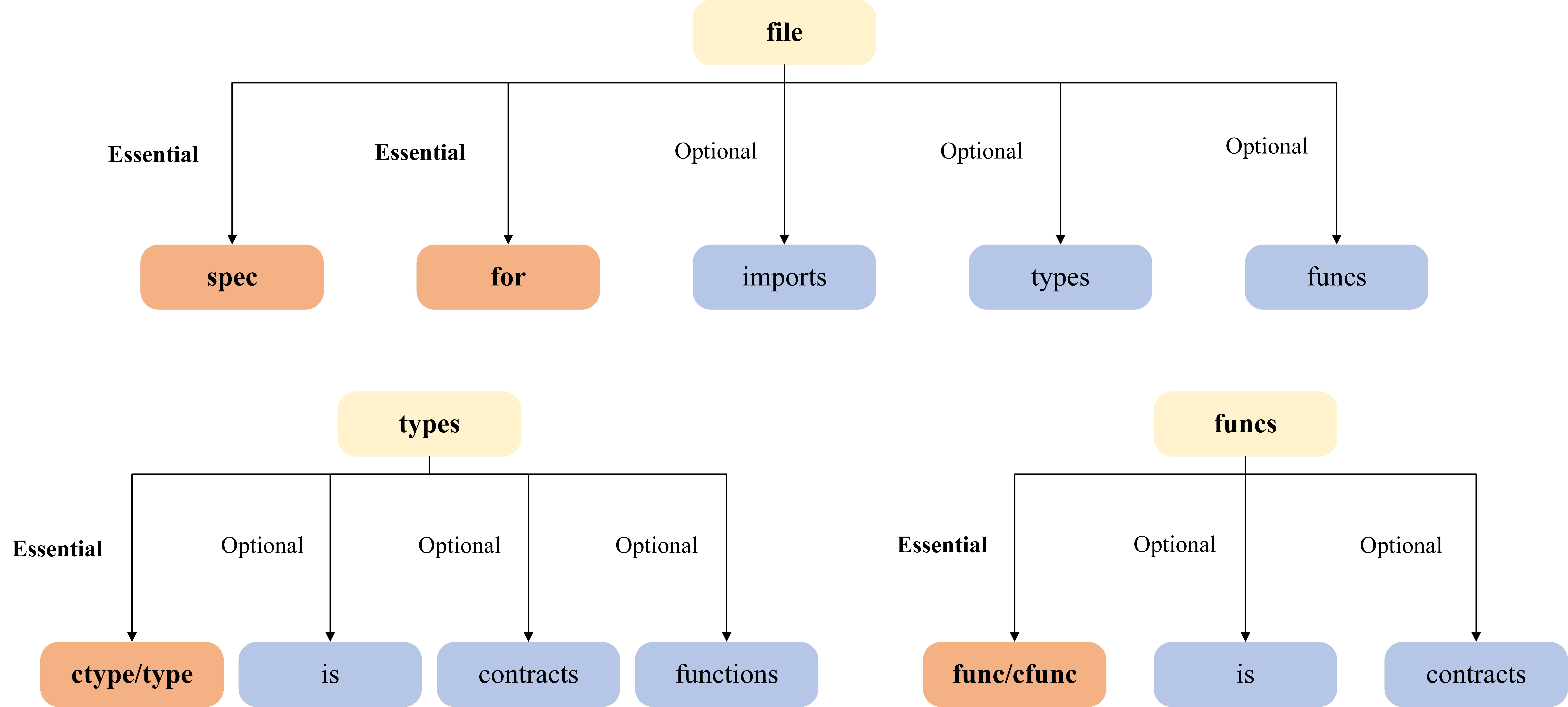}
    \caption{Composition rules of YAML nodes in an ASL file.}
    \label{fig2}
\end{figure}

Figure \ref{fig2} presents a schematic diagram of the composition of an ASL file. The keyword within the rounded rectangle denotes a YAML node, with its value corresponding to a specific entity type. The line signifies the relationship between nodes, with the element at the arrow's terminal being subordinate to the element at its origin.

As depicted in Figure \ref{fig2}, a complete specification file in the Agile Formal Specification Language consists of the following five YAML nodes: \texttt{spec}, \texttt{for}, \texttt{imports}, \texttt{types}, and \texttt{funcs}.

\noindent\textbf{"spec"}. Node \texttt{spec} is a essential node used to represent the specification identifier for easy reference. The value of this node is equal to the all-capital form of the specification file name.

\noindent\textbf{"for"}. Node \texttt{for} is a essential node used to determine the verification target file, which is the code file to be verified.

\noindent\textbf{"imports"}. Node \texttt{imports} is an optional node used to specify the files that must be incorporated. It enables the importation of either an ASL (.yaml) file or a K Framework (.k) file to facilitate verification. These imported files are utilized as components within the verification environment.

\noindent\textbf{"types"}. Node \texttt{types} is optional and used to describe types. These types follow the "sort as predicate" paradigm \cite{kaneiwa2004completeness}. There are four sub-nodes under \texttt{type}. One is essential: \texttt{type/ctype}. The other three are optional: \texttt{is}, \texttt{functions}, and \texttt{constructs}. Here are the functions of these sub-nodes:

\begin{itemize}
    \item Sub-node \texttt{type/ctype} is used to declare the type. When the type is used to assist in verification, it is represented by the keyword \texttt{type}. When it directly involves the verification target, it is represented by \texttt{ctype}. In this article, C language is employed as the verification target, hence it is identified by the keyword \texttt{ctype}. 
    \item Sub-node \texttt{is} is used to declare a type constructor. The type constructor means: if \texttt{t1 is t2}, then \texttt{t2} is the parent class of \texttt{t1}, and \texttt{t1} can use the function of \texttt{t2}; \texttt{t1} needs to satisfy the specification of \texttt{t2}.
    \item Sub-node \texttt{functions} is used to declare function definitions and can contain multiple function definition expressions.
    \item Sub-node \texttt{constructs} is used to declare specifications, and its value is an expression that can verify whether the value of the current type meets specific conditions.
\end{itemize}

\noindent\textbf{"funcs"}. Node \texttt{funcs} is an optional node used to describe functions. It has three sub-nodes. One is essential: \texttt{func/cfunc}. The other two are optional: \texttt{is} and \texttt{contracts}. Here are the functions of these sub-nodes:

\begin{itemize}
    \item Sub-node \texttt{func/cfunc} is used to declare a function. The function is represented by \texttt{func} when it is used to assist in verification; it is represented by \texttt{cfunc} when the function directly involves the verification target. Within this node, the input and output constraints of the function can be specified.
    \item Sub-node \texttt{is} is used to declare a function constructor. The function constructor means: if \texttt{f1 is f2}, then f1 may represent an equivalent, refinement, or expansion of \texttt{f2}. Equivalence is a special case of refinement or expansion, where \texttt{f1} and \texttt{f2} are completely equal. Refinement entails making a function more specific, such that the refined function must meet all the requirements of the original function but may impose additional constraints on the domain or value range. Expansion, on the other hand, introduces additional inputs and outputs beyond the original ones, without affecting the original inputs and outputs. \texttt{f1} should satisfy the specifications met by \texttt{f2}, thus enabling the reuse of specifications.
    \item Sub-node \texttt{contracts} is used to declare the specification of the function and can verify whether the value of the current function meets specific conditions.
\end{itemize}

In this context, we emphasize the key agile design aspects of ASL. Type-related functions and their definitions are collocated, similar to the design in object-oriented languages like Java. This arrangement facilitates a more intuitive reuse of functions alongside types. Moreover, specifications can be directly added to types. This allows the reuse of type specifications while using the types, enhancing the degree of specification reuse. Consequently, it eliminates the need for redundant specification writing. The separation of definitions from other specifications also enhances the agility of the specification-writing process.

\subsection{Example}
This subsection provides an application example of ASL to illustrate how the aforementioned design is applied in practice.

The code in Listing \ref{list1} is an agile formal specification in the example.yaml file, which implements the verification of the insertion function for a specific type of binary tree. It includes nodes such as \texttt{spec}, \texttt{imports}, \texttt{for}, \texttt{types}, and \texttt{funcs}.
\vspace{0.1cm}
\begin{lstlisting}[caption={An ASL example.},numbers=none,label={list1}]
  1 spec:EXAMPLE
  2 imports:c-verifier.k
  3 for:bst.c
  4 types:
  5  -type:HTree
  6   is:BinaryTree<htree(V::Int,Height::Int)>
  7     functions:
  8      IntSet tree_keys(HTree):
  9       -tree_keys(node(htree(V,Height),T0,T1)) = {V} U (tree_keys(T0) U tree_keys(T1))
 10       -tree_keys(leaf) => .IntSet
 11 funcs:
 12  -cfunc:insert
 13   inputs:HTree(T1) andBool min(t(.Set,int)) < Int V:Int andBool V:Int <= Int max(t(.Set,int))
 14  contracts:Htree(?T2) andBool tree_keys(?T2) == K{V} U tree_keys(T1)
 
\end{lstlisting}

\vspace{0.2cm}

\noindent\textbf{Line 1}. The \texttt{spec} node specifies that the specification is identified as \texttt{EXAMPLE}. The specification identifier is equal to the uppercase version of the specification file name, which is \texttt{example.yaml}.

\noindent\textbf{Line 2}. The \texttt{imports} node specifies the required imported file is \texttt{c-verifier.k}, which is a C language verifier based on the K Framework. It defines some basic syntax for C language verification.

\noindent\textbf{Line 3}. The \texttt{for} node specifies that the verification target file is \texttt{bst.c}, which contains the C code implementation of the binary tree data structure and functions.

\noindent\textbf{Lines 4-6}. The \texttt{types} node contains a type declaration variable \texttt{Htree} used to assist in verification. In the \texttt{is} sub-node, a function constructor is declared as follows: \texttt{is:BinaryTree<htree(v::Int, height::Int)>}. \texttt{BinaryTree} (binary search tree) is the parent class, and \texttt{HTree} is a specific type of \texttt{BinaryTree}. Each node in \texttt{HTree} contains two integer variables: \texttt{V} and \texttt{Height}. Through this construction method, \texttt{HTree} can use the functions in \texttt{BinaryTree} and also needs to satisfy the specifications of \texttt{BinaryTree}, achieving reuse of types, functions, and specifications simultaneously.

\noindent\textbf{Lines 7-10}. The \texttt{types} node also defines a function named \texttt{tree\_keys} which outlines two rules: (1) The first rule specifies the method for extracting all keys from an internal node of an \texttt{HTree}. \texttt{node(htree(V,Height), T0, T1)} represents an internal node with three components: \texttt{htree} and two subtrees \texttt{T0} and \texttt{T1}. The \texttt{tree\_keys} function is designed to extract the keys of all nodes in the tree. The rule declares that to extract keys from an internal node, the value \texttt{V} of the current node is first added to a set, followed by a recursive call to the \texttt{tree\_keys} function to extract the keys from the left and right subtrees \texttt{T0} and \texttt{T1}, respectively, and then merge them into this set.
(2) The second rule defines the process for extracting keys from the leaf nodes of an \texttt{HTree}. The rule implies that when extracting keys from a leaf node, the result is an empty set. In essence, these two rules detail the approach for recursively traversing an \texttt{HTree} and extracting keys from each node, ultimately aggregating these keys into a set. The \texttt{tree\_keys} function can serve as an auxiliary function during the verification of the insertion function.

\noindent\textbf{Lines 11-13}. The \texttt{funcs} node contains a function declaration variable \texttt{insert} that involves the verification target. It specifies that the verification target is the \texttt{insert} function in the \texttt{bst.c} file. The input of this function is constrained in the \texttt{inputs} node: \texttt{Htree(T1)} is used to check if \texttt{T1} satisfies the properties of a binary search tree. The constraint \texttt{min(t(.Set,int)) <= Int V:Int and Bool V:Int <= Int max(t(.Set,int))} requires that each value \texttt{V} in the tree must be within the range of the minimum and maximum values of the tree.

\noindent\textbf{Line 14}. The \texttt{contracts} section defines the specification for verification function: \texttt{Htree(?T2) andBool tree\_keys(?T2) == K\{V\} U tree\_keys(T1)}. This specification states that if \texttt{T2} is an \texttt{Htree} (\texttt{?} denotes an unmatched variable that will be matched or deduced during verification), and all the keys in \texttt{T2} are equal to the set formed by all the keys in \texttt{T1} plus an additional key \texttt{V}, then the specification is satisfied. This specification serves to validate whether the \texttt{insert} function correctly implements its intended functionality. Since the \texttt{tree\_keys} function from the type \texttt{Htree} is invoked here, it also needs to adhere to the specification of \texttt{BinaryTree}.

\subsection{Specification Agility Statement}
This subsection will elucidate the agile design philosophy integrated into the design of the Agile Formal Specification Language.

\begin{figure}
    \centering
    \includegraphics[width=\textwidth]{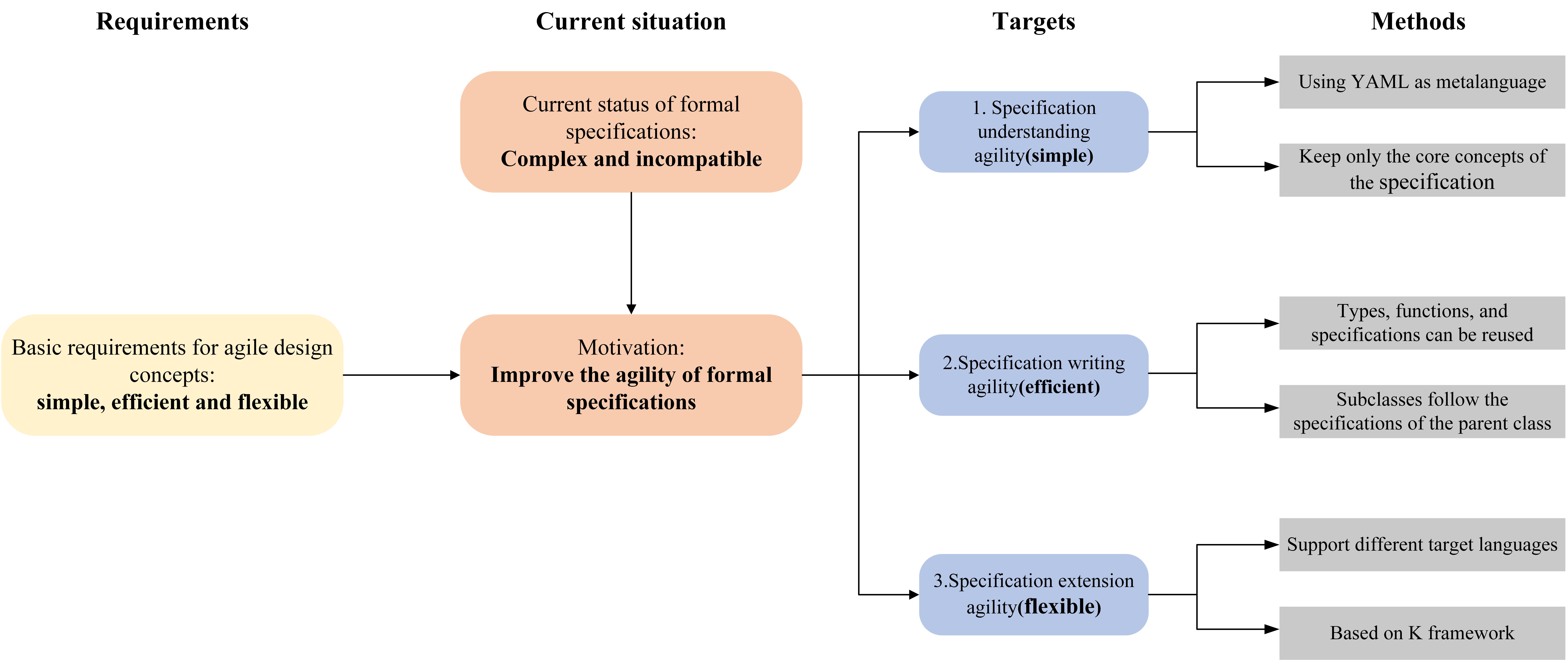}
    \caption{Agile design concepts in ASL.}
    \label{fig3}
\end{figure}

As illustrated in Figure \ref{fig3}, the fundamental requirements of agile design are simplicity, efficiency, and flexibility \cite{fowler2001agile}. To incorporate this philosophy into formal specifications, ASL follows a design principle. This principle aims to simplify specifications, enhance the efficiency of specification writing, and ensure the scalability of the specifications.

\noindent\textbf{Simplification of Specifications}. ASL utilizes YAML as its metalanguage, capitalizing on its ease of readability and structured nature. To further streamline the specification writing process, ASL retains only the essential core concepts of formal specifications.

\noindent\textbf{Enhancing Specification Writing Efficiency}. ASL allows for the reuse of types, functions, and specifications. Auxiliary functions necessary for verification can be defined using the \texttt{func} or \texttt{type} nodes. These definitions can be based on existing functions or types, facilitating reuse or extension and avoiding redundant writing. Only specific specifications need to be incrementally added.

\noindent\textbf{Scalability of Specifications}. ASL supports different target languages. In this paper's introduction, the verification target is C language. It can also import K Framework verifiers for different languages through the \texttt{imports} node to support the verification of other languages \cite{stefuanescu2016semantics}. This takes advantage of the K Framework's ease of extension.

\section{Specification Translation Algorithm}\label{sec4}
This section presents the design of an algorithm for transforming agile formal specifications into K formal specifications.

\subsection{Overall Process}
\begin{figure}
    \centering
    \includegraphics[width=0.6\textwidth]{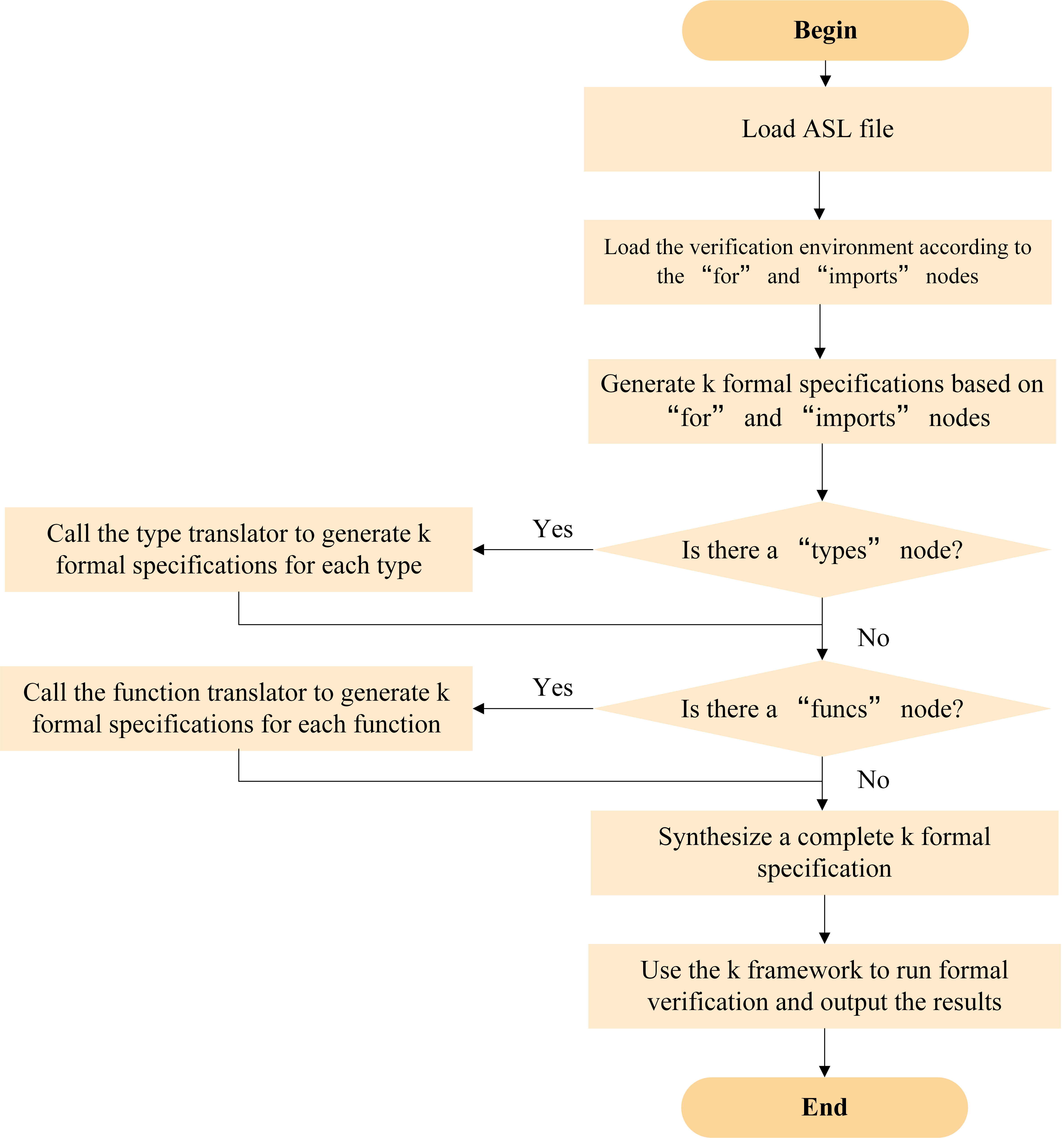}
    \caption{Translation algorithm flow chart.}
    \label{fig4}
\end{figure}

Figure \ref{fig4} illustrates the entire process of ASL verification. It starts by loading the ASL file, followed by loading the verification environment based on the specification target specified in the \texttt{for} node and the imported files listed in the \texttt{imports} node. This process generates partial K formal specifications. Subsequently, the type and function translation modules generate corresponding K formal specifications for types and functions, respectively. These specifications are then combined to form a complete K formal specification. Finally, the formal verification is executed using commands from the K Framework, and the verification results are output. The type and function translation modules will be introduced in the next subsection.

\subsection{Translator Module}\label{subsec4.2}
This subsection describes the functions of the translator module.

\noindent\textbf{Type Translator Module}. The role of the type translator is to convert the values of the \texttt{types} node in ASL (represented by the variable \texttt{ASL\_type}) into the form of K formal specifications (represented by the variable \texttt{k\_type}). The detailed steps can be found in Algorithm \ref{A1}.

\begin{algorithm}
  \caption{Type translator module working steps}
  \KwIn{\texttt{ASL\_type}}
  \KwOut{ \texttt{k\_type}}
  \small
  \SetKwProg{Fn}{Procedure}{:}{}
  \SetKwFunction{FMain}{main}
  \Fn{initialization()}{
    Initialize \texttt{k\_type} = empty string\;
  }
  \Fn{handle\_type\_node()}{
    \If{\texttt{type} exists in \texttt{ASL\_type}}{
      parse \texttt{ASL\_type[type]} and add the result to \texttt{k\_type}\;
    }
    \If{\texttt{ctype} exists in \texttt{ASL\_type}}{
      parse \texttt{ASL\_type[ctype]} and add the result to \texttt{k\_type}\;
    }
    \If{Neither \texttt{type} nor \texttt{ctype} exists in \texttt{ASL\_type}}{
      Report an error and exit the program\;
    }
  }

  \Fn{handle\_is\_node()}{
    \If{\texttt{is} exists in \texttt{ASL\_type}}{
      \If{the value of \texttt{ASL\_type[is]} is of list type}{
        \For{each value(constructor) in \texttt{ASL\_type[is]}}{
        Parse each constructor to obtain the definition, functions and specifications of the parent class\;
        Add the parsing results to \texttt{k\_type} by splicing strings\;
        }
      }
      
      \ElseIf{the value of \texttt{ASL\_type[is]} is of string type}{
        Perform lexical and syntactic analysis of the value (constructor) and obtain the definition, functions and specifications of the parent class\;
        Add the parsing results to \texttt{k\_type}\;
      }
    }
  }

  \Fn{handle\_functions\_node()}{
    \If{\texttt{functions} exists in \texttt{ASL\_type}}{
      \For{each value(function rules) in \texttt{ASL\_type[functions]}}{
      Add function rules to \texttt{k\_type}\;
      }
    }
  }

  \Fn{handle\_contracts\_node()}{
    \If{\texttt{contracts} exists in \texttt{ASL\_type}}{
      \If{the value of \texttt{ASL\_type[contracts]} is a list type}{
        \For{each value(specification) in \texttt{ASL\_type[contracts]}}{
        Splice each specification with \texttt{andBool} and add it to \texttt{k\_type}\;
        }
      }
      \ElseIf{the value of \texttt{ASL\_type[contracts]} is of string type}{
        Add value (specification) to \texttt{k\_type}\;
      }
    }
  }
  \Fn{main()}{
    initialization()\;
    handle\_type\_node()\;
    handle\_is\_node()\;
    handle\_functions\_node()\;
    handle\_contracts\_node()\;
    Add newline character to \texttt{k\_type}\;
    return \texttt{k\_type}\;
  }
  \label{A1}
\end{algorithm}

\noindent\textbf{Function Translator Module}. The function translator is responsible for converting the function nodes in ASL (represented by the variable \texttt{ASL\_func}) into the form of k formal specifications (represented by the variable \texttt{k\_func}). The implementation steps are similar to those of the type translator and will not be repeated here.

\section{Evaluation}\label{sec5}
In this section, the Agile Formal Specification Language is used to perform specification writing and verification experiments on C code cases. Then it is compared with existing formal specification methods to test the role of this method in improving the agility of specification writing.

\subsection{Comparison Method}
In the experiments, this paper's proposed Agile Formal Specification Language and existing formal verification tools were used to write specifications and validate C code cases, with a comparative analysis conducted. 

The K Framework \cite{rosu2017k} and CBMC\cite{kroening2014cbmc} were selected as the comparison benchmarks. The K Framework serves as the foundation for the Agile Formal Specification Language, while CBMC is a widely-used tool for verifying C code.

Due to the absence of quantitative metrics for agility in the field of formal verification and the lack of a fixed method for writing formal specifications, this paper employs the following comparative approach:

\begin{itemize}
    \item When compared with the K Framework, the complexity of the specification can be evaluated by comparing the number of code lines. This comparative method is effective because the expression syntax of ASL is identical to that of the K Framework. The selection of cases and the method for writing K formal specifications drew upon the work of Stefănescu et al.
\cite{stefuanescu2016semantics}.
    \item When compared with CBMC, since the syntax is different and there is no standard way to write specifications, the code lines cannot be used as a comparative metric. Therefore, this paper examines the differences in agility between two specification languages through the analysis of a case study.
\end{itemize}

During the specification translation and verification process, Python 3.8 was used to implement the specification translation algorithm, and the K Framework 4.0 was employed to run the verification.

\subsection{Results and Analysis}
\begin{figure}
    \centering
    \includegraphics[width=\textwidth]{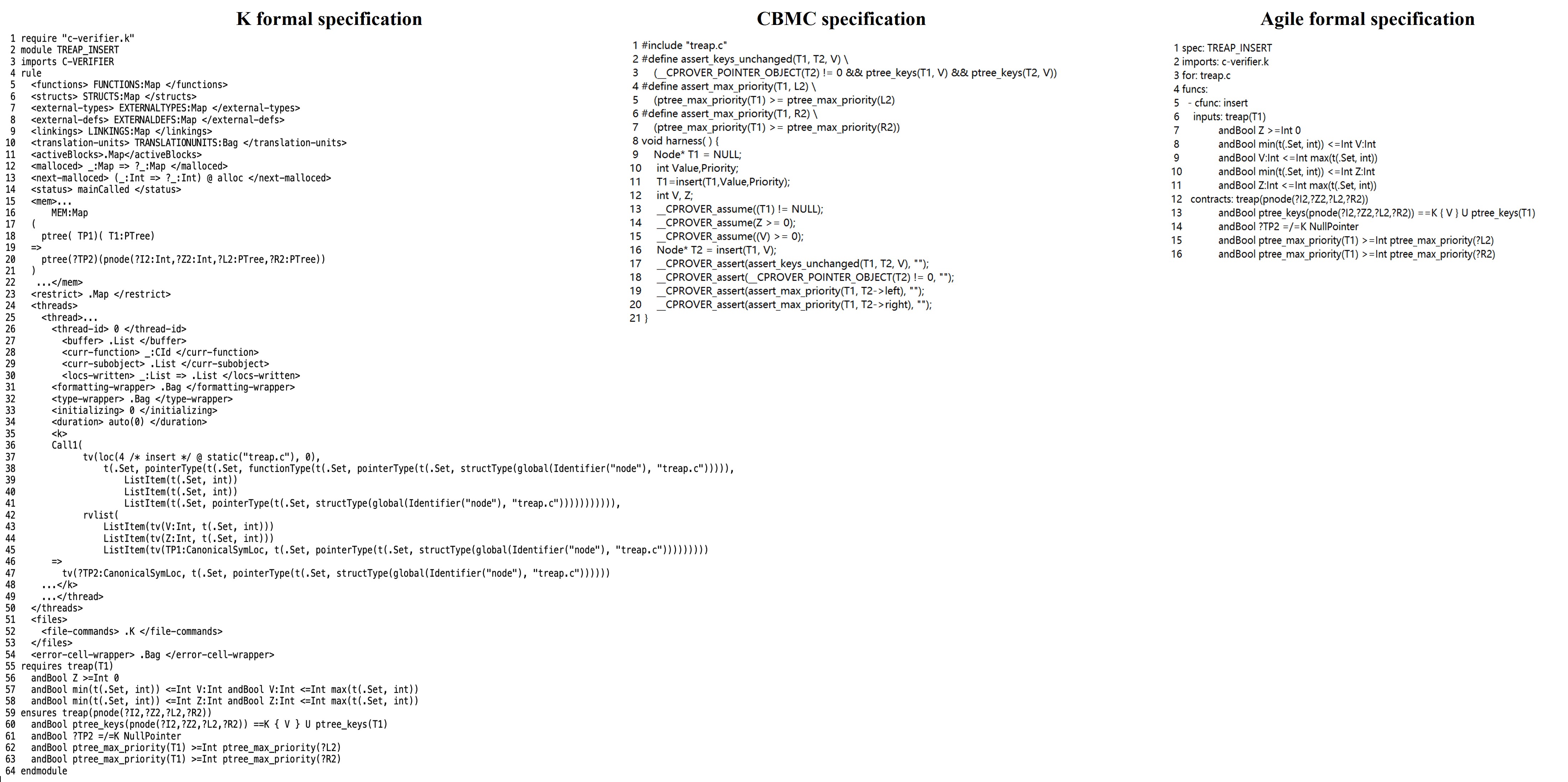}
    \caption{Specification comparison for 3 kinds of specification languages.}
    \label{fig5}
\end{figure}

Figure \ref{fig5} illustrates a comparison of verification specifications for the Treap insertion function \cite{lammich2019priority}, using K formal specification, CBMC specification, and ASL. The K formal specification is produced by the specification translation algorithm. 

From Figure \ref{fig5}, it is evident that ASL can describe specifications more concisely than K formal specifications, using only 16 lines. When compared to CBMC specifications, although the number of lines is roughly the same, ASL's structure is clearer and more readable. This is because ASL uses YAML nodes for structured representation, dividing the specification into parts. This structure makes it easier to modify and extend specifications. In contrast, CBMC specifications require specific primitives like \texttt{\_\_CPROVER\_assert} for verification at precise program locations. This process can be complex.

Table \ref{table1} presents a comparison of code lines between ASL and K formal specifications for other cases.

\begin{table}
\centering
\caption{Line numbers comparison for Treap cases.}\
\begin{tabular}{|l|l|l|l|l|}
\hline
Data Structure & Function & Code Lines(ASL) & Code Lines(K) & Reduction Ratio(\%)\\
\hline
Treap & Insert Node & 16 & 64 & 75.00\% \\
Treap & Find Node  & 9 & 48 & 81.25\%\\
Treap & Delete Node & 18 & 111 & 83.78\% \\
\hline
\end{tabular}
\label{table1}
\end{table}

As shown in Table \ref{table1}, for the Treap’s insertion function, ASL achieves a 75\% reduction in code lines compared to K formal specifications. In the case of the search function, this reduction is 81.25\%, and for the deletion function, it is 83.78\%.

To further validate the effectiveness of ASL, this paper also performed verification and comparison on additional cases, with the experimental results summarized in Table \ref{table2}.

\vspace{-1cm}

\begin{table}
\centering
\caption{Line numbers comparison for other cases.}
\begin{tabular}{|l|l|l|l|l|}
\hline
Data Structure & Function & Code Lines(ASL) & Code Lines(K) & Reduction Ratio(\%) \\
\hline
Linked List & Reverse List & 69 & 294 & 76.53\%\\
 & Append Node & 64 & 265 & 75.85\%\\
 & Bubble Sort & 96 & 473 & 79.70\%\\
 & Insertion Sort & 101 & 501 & 79.84\%\\
 & Merge Sort & 103 & 483 & 78.67\%\\
 & Quick Sort & 107 & 548 & 80.47\%\\
\hline
Binary Search Tree & Insert Node & 13 & 126 & 89.68\%\\
 & Search Node & 29 & 69 & 57.97\%\\
 & Delete Node & 118 & 308 & 61.69\%\\
\hline
Red-Black Tree & Insert Node & 38 & 170 & 77.65\%\\
 & Search Node & 8 & 47 & 82.98\%\\
 & Delete Node & 142 & 427 & 66.74\%\\
\hline
\end{tabular}
\label{table2}
\end{table}

As shown in Table \ref{table2}, ASL can significantly reduce the workload for specification writing in verification cases of data structures. These structures include linked lists, binary search trees, and red-black trees. On average, ASL achieves a 76.07\% reduction in code lines compared to K formal specifications.

From the experiments conducted, it can be concluded that the ASL aligns with the agility demands of formal specifications, boasting simplicity, efficiency, and scalability. ASL can produce complex K formal specifications using the specification translation algorithm, which lessens the complexity and effort involved in writing specifications. In contrast to CBMC, ASL offers a more structured and clearer syntax for specification depiction.

\section{Conclusion and Future Work}\label{sec6}
To resolve the complexities associated with writing specifications in the formal verification process, this paper proposes an Agile Formal Specification Language based on the K Framework. Firstly, the design of the ASL integrates agile design principles into the formal specification, improving the readability, efficiency, and scalability of specification writing. Subsequently, a specification translation algorithm is employed to convert ASL into K formal specifications to support verification. Experimental results demonstrate that the proposed method improves the agility of specification writing compared to existing formal specification languages such as K Framework and CBMC.

However, the current Agile Formal Specification Language has limitations. Verification requires conversion to K formal specifications, which can limit case selection and sometimes cause syntax errors. Our next goal is to research how to enable ASL to directly and efficiently use solvers for verification, reducing reliance on the K Framework. Additionally, we plan to apply ASL in more software development projects and continue improving its reliability, completeness, and portability.

%
%

%
%
%
\bibliographystyle{splncs04}
\bibliography{mybibliography}
%





\end{document}